\documentclass[12pt]{article}

\normalbaselines
\begin{document}

\centerline{\bf NON-MINIMAL EINSTEIN-YANG-MILLS-DILATON THEORY }

\vskip 5 mm

\centerline{\bf Alexander Balakin\footnote{e-mail:
Alexander.Balakin@ksu.ru} }

\vskip 0.3cm

\centerline{\it Kazan State University,}

\centerline{\it Kremlevskaya str., 18, 420008, Kazan, Russia}

\vskip 0.3cm

\centerline{\bf Heinz Dehnen\footnote{e-mail:
Heinz.Dehnen@uni-konstanz.de} } \vskip 0.3cm

\centerline{\it Universit\"at Konstanz, Fachbereich Physik,}
\centerline{\it Fach M677, D-78457, Konstanz, Germany}

\vskip 0.3cm

\date{\today}

\begin{abstract}

We establish a new non-minimal Einstein-Yang-Mills-dilaton model,
for which the Lagran\-gian is linear in the curvature and contains
eight arbitrary functions of the scalar (dilaton) field. The
self-consistent system of equations for the non-minimally coupled
gauge, scalar and gravitational fields is derived. As an example
of an application we discuss the model with pp-wave symmetry. Two
exact explicit regular solutions of the whole system of master
equations, belonging to the class of pp-wave solutions, are
presented.

\end{abstract}

\section{Introduction}\label{Intro}

The Einstein-Maxwell-dilaton theory and its non-Abelian
generalization, Einstein-Yang-Mills-dilaton (EYMd) theory, attract
serious attention, since they have a supplementary, dilatonic,
degree of freedom for the structure modeling of the
gravitationally coupled systems. A lot of impressive results are
obtained in the framework of the {\it minimal} EYMd theory in the
cosmological and string contexts, as well as in the application to
the colored static spherically and axially symmetric objects (see,
e.g., \cite{1} - \cite{5} and references therein). A new impetus
to the development of the EYMd theory has been given by the
discovery of the accelerated expansion of the Universe. Numerous
attempts have been made to consider modified theories of
gravitational interaction, such as $F(R)$-gravity,
Gauss-Bonnet-gravity, etc., as alternatives to the dark energy
(see, e.g., \cite{6} and references therein). One of the
directions in such a generalization of the Einstein theory of
gravity is connected with a {\it non-minimal} extension of the
field theory. Some historical details, review and references,
related to the non-minimal interaction of gravity with scalar and
electromagnetic fields, can be found, e.g., in \cite{FaraR}. As
for the non-minimal generalization of the Einstein-Yang-Mills
theory, there are two different approaches. The first one is based
on the dimensional reduction of the Gauss-Bonnet action \cite{MH},
the alternative way is connected with the non-Abelian
generalization of the non-minimal Einstein-Maxwell theory
\cite{BZ,BDZ} along the lines proposed by Drummond and Hathrell
for the linear electrodynamics \cite{Drum}. Recently in
\cite{Bamba} one of the variants of the non-linear non-minimal
generalizations of the Einstein-Yang-Mills theory has been
developed, which combines, in fact, two ideas: the idea of
$F(R)$-gravity, on the one hand, and the idea of Ricci-dilaton, on
the other hand. The paper \cite{Bamba} has stimulated in some
respects the preparation of this note.

\section{Fundamentals of the non-minimal EYMd model}\label{sec1}

\subsection{General concepts and definitions}

We consider a non-minimal Einstein-Yang-Mills-dilaton model
quadratic in the Yang-Mills field strength $F_{ik}^{(a)}$,
quadratic in the derivative of the scalar (dilaton) field
$\nabla_k \Phi$, and linear in the curvature. Such a model can be
described self-consistently using the action functional of the
type
\begin{eqnarray}\label{0act}
S_{({\rm EYMd})} = \int d^4 x \sqrt{-g}\ \left\{ \frac{R {+} 2
\Lambda}{\kappa}{+}\frac{1}{2} C^{ikmn}_{(a)(b)}(\Phi) \
F^{(a)}_{ik} F_{mn}^{(b)} {-}{\cal C}^{ik}(\Phi) \nabla_i\Phi
\nabla_k \Phi {+} {\cal F}(\Phi) R {+} V(\Phi) \right\}\,,
\end{eqnarray}
which is, on the one hand, a dilatonic extension of the
non-minimal Einstein-Yang-Mills model \cite{BZ}, and, on the
other hand, a reduction of the Einstein-Yang-Mills-Higgs
functional \cite{BDZ} to the case, when the $SU(N)$ Higgs
miltiplet $\Phi^{(a)}$ degenerates into the scalar singlet $\Phi$.

The used definitions are standard:  $g = {\rm det}(g_{ik})$ is the
determinant of a metric tensor $g_{ik}$, $R$ is the Ricci scalar,
$\Lambda$ is the cosmological constant, the multiplet of real
tensor fields $F^{(a)}_{ik}$ describes the strength of gauge
field, the symbol $\Phi$ denotes the real scalar field, associated
with a dilaton, $\nabla_k$ denotes the covariant derivative, $V(
\Phi)$ is a potential of the scalar field. Latin indices without
parentheses run from 0 to 3, $(a)$ and $(b)$ are the group
indices, running from $(1)$ to $(N^2-1)$ for the model with
$SU(N)$ symmetry. The quantities $C^{ikmn}_{(a)(b)}$ and ${\cal
C}^{ik}$ denote the so-called constitutive tensors for the gauge
and scalar fields, respectively. They contain neither Yang-Mills
strength tensor $F^{(a)}_{ik}$, nor the derivative of the scalar
field. Depending on the model under consideration, these tensors
can be constructed using space-time metric, its first derivatives
(through the covariant derivative $\nabla_k$), second derivatives
(through the Riemann tensor $R^{i}_{\ klm}$, Ricci tensor $R_{km}$
and Ricci scalar $R$), etc. In addition, the scalar field $\Phi$,
time-like velocity four-vector $U^k$ and space-like director(s)
$N^k$, and their covariant derivatives, $\nabla_k U_l$ and
$\nabla_k N_l$, can be constructive elements of the constitutive
tensors.

We follow the definitions of the book \cite{Rubakov} and consider
the Yang-Mills field ${\bf F}_{mn}$ taking values in the Lie
algebra of the gauge group $SU(N)$ (adjoint representation):
\begin{equation}
{\bf F}_{mn} = - i {\cal G} {\bf t}_{(a)} F^{(a)}_{mn} \,, \quad
{\bf A}_m = - i {\cal G} {\bf t}_{(a)} A^{(a)}_m  \,.
\label{represent}
\end{equation}
The generators ${\bf t}_{(a)}$ are Hermitian and traceless. The
symmetric tensor $G_{(a)(b)}$ defined as
\begin{equation}
G_{(a)(b)} \equiv 2 {\rm Tr} \ {\bf t}_{(a)} {\bf t}_{(b)} \,,
\label{scalarproduct}
\end{equation}
plays a role of a metric in the group space. The tensor
$F^{(a)}_{mn}$ is connected with the potentials of the gauge field
$A^{(a)}_i$ by the formulas \cite{Rubakov,SDObook}
\begin{equation}
F^{(a)}_{mn} = \nabla_m A^{(a)}_n - \nabla_n A^{(a)}_m + {\cal G}
f^{(a)}_{\cdot (b)(c)} A^{(b)}_m A^{(c)}_n \,. \label{Fmn}
\end{equation}
The symbols $f^{(a)}_{\cdot (b)(c)}$ denote the real structure
constants of the gauge group $SU(N)$. The tensor $F^{ik}_{(a)}$
satisfies the relation
\begin{equation}
\hat{D}_k {}^*\! F^{ik}_{(a)} \equiv \nabla_k {}^*\! F^{ik}_{(a)}
- {\cal G} f^{(c)}_{\cdot (b)(a)} A^{(b)}_m {}^*\! F^{ik}_{(c)} =
0 \,. \label{Aeq2}
\end{equation}
The symbol $\hat{D}_k$ denotes the gauge-covariant derivative.
For the derivative of arbitrary tensor defined in the group space
we use the following rule \cite{Akhiezer}:
\begin{eqnarray}
\hat{D}_m Q^{(a) \cdot \cdot \cdot}_{\cdot \cdot \cdot (d)} \equiv
\nabla_m Q^{(a) \cdot \cdot \cdot}_{\cdot \cdot \cdot (d)} + {\cal
G} f^{(a)}_{\cdot (b)(c)} A^{(b)}_m Q^{(c) \cdot \cdot
\cdot}_{\cdot \cdot \cdot (d)} - {\cal G} f^{(c)}_{\cdot (b)(d)}
A^{(b)}_m Q^{(a) \cdot \cdot \cdot}_{\cdot \cdot \cdot (c)} +...
\,. \label{DQ2}
\end{eqnarray}
The asterisk relates to the dual tensor
\begin{equation}
{}^*\! F^{ik}_{(a)} = \frac{1}{2}\epsilon^{ikls} F_{ls (a)} \,,
\label{dual}
\end{equation}
where $\epsilon^{ikls} = \frac{1}{\sqrt{-g}} E^{ikls}$ is the
Levi-Civita tensor, $E^{ikls}$ is the completely antisymmetric
symbol with $E^{0123} = - E_{0123} = 1$.

\subsection{Non-minimal EYMd model based on eight arbitrary functions of the dilaton field}

In the papers \cite{BZ,BDZ} we focused on the three-, five-, six-
and seven-{\it parameter} models, considering the parameters
$q_1$, $q_2$, etc., as phenomenological non-minimal coupling
constants. Now, following the main idea of dilatonic extension of
the Einstein-Maxwell and Einstein-Yang-Mills theories (see, e.g.,
\cite{Fort}), we assume that the (non-minimal) constitutive
tensors $ C^{ikmn}_{(a)(b)}$ and ${\cal C}^{ik}$ contain
arbitrary functions of the dilaton field $\Phi$ instead of
coupling constants. Our ansatz for the action functional of the
non-minimal EYMd model is the following
$$
S_{({\rm NMEYMd})} = \int d^4 x \sqrt{-g}\ \left\{ \frac{R + 2
\Lambda}{\kappa}+ \frac{1}{2} f_0(\Phi)F^{(a)}_{ik} F^{ik}_{(a)} -
{\cal F}_0(\Phi) \nabla_m\Phi \nabla^m\Phi + V(\Phi) + \right.
$$
\begin{equation}\label{1act}
\left.  + {\cal F}(\Phi) R  +\frac{1}{2} {\cal R}^{ikmn}(\Phi)
F^{(a)}_{ik} F_{mn(a)} - {\Re}^{\,mn}(\Phi) \nabla_m \Phi \nabla_n
\Phi \right\}\,.
\end{equation}
The function $f_0(\Phi)$ describes a multiplier of the
dilaton-type, which is traditional for the {\it minimal} theory;
it is placed in front of the first invariant $F^{(a)}_{ik}
F^{ik}_{(a)}$ of the gauge field and is considered to be positive.
In analogy with $f_0(\Phi)$ we introduce the multiplier ${\cal
F}_0(\Phi)$ in front of the invariant $\nabla_m\Phi \nabla^m\Phi$.
The function ${\cal F}(\Phi)$ is typical for the non-minimal
extension of the scalar field theory; traditionally, one uses the
term ${\cal F}(\Phi) R$ in the well-known form $\xi R \Phi^2$. The
so-called susceptibility tensor ${\cal R}^{ikmn}$
$$
{\cal R}^{ikmn}(\Phi) \equiv \frac{1}{2} f_1(\Phi)
R\,(g^{im}g^{kn}-g^{in}g^{km}) + f_3(\Phi) R^{ikmn} +
$$
\begin{equation}
+ \frac{1}{2} f_2(\Phi)(R^{im}g^{kn} - R^{in}g^{km} + R^{kn}g^{im}
-R^{km}g^{in}) \,, \label{sus}
\end{equation}
contains three arbitrary functions $f_1(\Phi)$, $f_2(\Phi)$,
$f_3(\Phi)$ instead of curvature coupling parameters $q_1$,
$q_2$, $q_3$ (compare with \cite{BZ,BDZ,BL}). The tensor
$\Re^{\,mn}$
\begin{equation}\label{Reqq}
\Re^{\,mn}(\Phi) \equiv {f_4(\Phi)}R g^{mn} + f_5(\Phi) R^{mn}\,,
\end{equation}
describes the so-called derivative coupling of the scalar field to
the curvature (see, e.g., \cite{AC}), but now we consider
arbitrary functions $f_4(\Phi)$ and $f_5(\Phi)$ instead of
coupling parameters $q_4$ and $q_5$ (see \cite{BDZ}). Finally, it
is worth noting, that the constitutive tensors, introduced in
(\ref{0act}), have now the following form
\begin{equation}
C^{ikmn}_{(a)(b)}(\Phi) {=} \left[ \frac{1}{2} f_0(\Phi)( g^{im}
g^{kn} {-} g^{in} g^{km}) {+} {\cal R}^{ikmn}(\Phi) \right]
G_{(a)(b)} \,, \quad {\cal C}^{mn}(\Phi) {=} \left[ g^{mn} {\cal
F}_0(\Phi) {+} \Re^{\,mn}(\Phi) \right] .
 \label{Hik11}
\end{equation}
These quantities are linear in the curvature and are arbitrary
functions of the dilaton field. They can be used for the
definition of dilatonically modified color permittivities, as well
as color and acoustic metrics along the lines described in
\cite{BDZ}.

\subsubsection{Non-minimal extension of the Yang-Mills equations}

The variation of the action $S_{({\rm NMEYMd})}$ with respect to
the Yang-Mills potential $A^{(a)}_i$ yields
\begin{equation}
\hat{D}_k {\cal H}^{ik}_{(a)}  =  0 \,,  \quad {\cal
H}^{ik}_{(a)} = C^{ikmn}_{(a)(b)} \ F^{(b)}_{mn} \,, \label{Heqs}
\end{equation}
where the term ${\cal H}^{ik}_{(a)}$ with constitutive tensor from
(\ref{Hik11}) describes a non-minimal color excitation in analogy
with electrodynamics of continuous media (see, e.g.,
\cite{HehlObukhov}). In this context the quantity ${\cal
R}^{ikmn}(\Phi) G_{(a)(b)}$ can be indicated as a color-dilatonic
susceptibility tensor.

\subsubsection{Non-minimal extension of the scalar field equations}

The variation of the action $S_{({\rm NMEYMd})}$ with respect to
the scalar $\Phi$ gives the master equation for the dilaton field
$$
\nabla_m \left\{ \left[ g^{mn}{\cal F}_0(\Phi) + \Re^{\,mn}(\Phi)
\right]\nabla_n \Phi \right\} = - \frac{1}{2}
\frac{d}{d\Phi}V(\Phi) - \frac{1}{2} R \ \frac{d}{d\Phi}{\cal
F}(\Phi)  + \frac{1}{2} \nabla_m \Phi \nabla^m \Phi
\frac{d}{d\Phi}{\cal F}_0(\Phi) -
$$
\begin{equation}\label{Heq}
- \frac{1}{4}\left[  F^{mn}_{(a)} F_{mn}^{(a)} \frac{d}{d\Phi}
f_0(\Phi) + F_{ik(a)} F^{(a)}_{mn} \ \frac{d}{d\Phi} {\cal
R}^{ikmn}(\Phi) \right] - \frac{1}{2}  \nabla_m \Phi \nabla_n
\Phi \ \frac{d}{d\Phi} \Re^{mn}(\Phi)\,.
\end{equation}
Clearly, this equation is coupled to the non-minimally extended
Yang-Mills equation (\ref{Heqs}), when the constitutive tensor
(\ref{Hik11}) is dilatonically extended.

\subsubsection{Master equations for the gravitational field}

The variation of the action $S_{({\rm NMEYMd})}$ with respect to
the metric gives the non-minimally extended equations of the
Einstein type
$$
\left(R_{ik}-\frac{1}{2}Rg_{ik}\right)\cdot\left[1+\kappa {\cal
F}(\Phi) \right] = \Lambda g_{ik} + \kappa \left(\nabla_i \nabla_k
- g_{ik}\nabla_m \nabla^m \right){\cal F}(\Phi) +
$$
\begin{equation}
+ \kappa \left[ T^{(YM)}_{ik} + T^{(\Phi)}_{ik} + T^{({\rm
NonMin})}_{ik} \right] \,. \label{Ein}
\end{equation}
Here the term $T^{(YM)}_{ik}$
\begin{equation}
T^{(YM)}_{ik} \equiv f_0(\Phi) \left[ \frac{1}{4} g_{ik}
F^{(a)}_{mn}F^{mn}_{(a)} - F^{(a)}_{in}F_{k\,(a)}^{\ n} \right]\,,
\label{TYM}
\end{equation}
is a stress-energy tensor of Yang-Mills field extended by the
dilatonic multiplier. The term $T^{(\Phi)}_{ik}$
\begin{equation}
T^{(\Phi)}_{ik}= {\cal F}_0(\Phi) \left[ \nabla_i \Phi \nabla_k
\Phi -\frac{1}{2}g_{ik}\nabla_m \Phi \nabla^m \Phi \right]+
\frac{1}{2} V(\Phi) g_{ik} \,,
\end{equation}
is a dilatonically extended stress-energy tensor of the scalar
field. The term $T^{({\rm NonMin})}_{ik}$ can be decomposed into
five parts
\begin{equation}
T^{({\rm NonMin})}_{ik} =   T^{(I)}_{ik} + T^{(II)}_{ik} +
T^{(III)}_{ik} + T^{(IV)}_{ik} + T^{(V)}_{ik} \,, \label{Tdecomp}
\end{equation}
which are enumerated corresponding to the functions $f_1$, $f_2$,
...$f_5$:

$$
T^{(I)}_{ik} {=}  f_1(\Phi) R \left[\frac{1}{4} g_{ik}
F^{(a)}_{mn}F^{mn}_{(a)} {-} F^{(a)}_{in}F_{k\,(a)}^{\ n} \right]
{-} \frac{1}{2} f_1(\Phi) R_{ik} F^{(a)}_{mn}F^{mn}_{(a)} {+}
$$
\begin{equation}
+ \frac{1}{2} \left[ {\hat{D}}_{i} {\hat{D}}_{k} {-} g_{ik}
{\hat{D}}^l {\hat{D}}_l \right] \left[f_1(\Phi)
F^{(a)}_{mn}F^{mn}_{(a)} \right] \,, \label{TI}
\end{equation}
$$
T^{(II)}_{ik} = -\frac{1}{2}g_{ik}\left\{{\hat{D}}_{m}
{\hat{D}}_{l}\left[f_2(\Phi) F^{mn(a)}F^{l}_{\ n(a)}\right]-
f_2(\Phi) R_{lm}F^{mn (a)} F^{l}_{\ n(a)} \right\}
$$
$$
- f_2(\Phi) F^{ln(a)} \left[R_{il}F_{kn(a)} +
R_{kl}F_{in(a)}\right]- f_2(\Phi) R^{mn}F^{(a)}_{im} F_{kn(a)} -
\frac{1}{2} {\hat{D}}^m{\hat{D}}_m \left[f_2(\Phi) F^{(a)}_{in}
F_{k\,(a)}^{\ n}\right]
$$
\begin{equation}%
+\frac{1}{2}{\hat{D}}_l \left\{ {\hat{D}}_i \left[ f_2(\Phi)
F^{(a)}_{kn}F^{ln}_{(a)} \right] + {\hat{D}}_k \left[f_2(\Phi)
F^{(a)}_{in}F^{ln}_{(a)} \right] \right\} \,, \label{TII}
\end{equation}%
$$
T^{(III)}_{ik} = \frac{1}{4}f_3(\Phi) g_{ik}
R^{mnls}F^{(a)}_{mn}F_{ls(a)}- \frac{3}{4} f_3(\Phi) F^{ls(a)}
\left[F_{i\,(a)}^{\ n} R_{knls} + F_{k\,(a)}^{\ n}R_{inls}\right]
$$
\begin{equation}%
-\frac{1}{2}{\hat{D}}_{m} {\hat{D}}_{n} \left\{f_3(\Phi) \
\left[F_{i}^{ \ n (a)}F_{k\,(a)}^{ \ m} + F_{k}^{ \ n(a)}
F_{i\,(a)}^{ \ m} \right]\right\} \,, \label{TIII}
\end{equation}%
$$
T^{(IV)}_{ik}= f_4(\Phi) \
\left[\left(R_{ik}-\frac{1}{2}Rg_{ik}\right)\nabla_m \Phi \nabla^m
\Phi + R\,\nabla_i \Phi \nabla_k \Phi \right]
$$
\begin{equation}\label{TIV}
+\left(g_{ik}\nabla_n \nabla^n-\nabla_i\nabla_k
\right)\left[f_4(\Phi) \ \nabla_m \Phi \nabla^m \Phi \right]\,,
\end{equation}
$$
T^{(V)}_{ik}= f_5(\Phi) \ \nabla_m \Phi \left[R_i^m \nabla_k \Phi
+ R_k^m \nabla_i \Phi \right] +
$$
$$
+\frac{1}{2}g_{ik}\left[ \nabla_m \nabla_n - R_{mn} \right]
\left[f_5(\Phi) \ \nabla^m \Phi \nabla^n \Phi \right]
$$
\begin{equation}\label{TV}
-\frac{1}{2}\,\nabla^m\biggl\{\nabla_i\left[f_5(\Phi) \
\nabla_m\Phi \nabla_k\Phi \right]+
    \nabla_k \left[f_5(\Phi) \ \nabla_m\Phi \nabla_i\Phi \right]-
    \nabla_m\left[f_5(\Phi) \ \nabla_i \Phi \nabla_k \Phi \right]\biggr\}\,.
\end{equation}
Straightforward calculations, based on the Bianchi identities and
on the properties of the Riemann tensor, show that the equality
\begin{equation}
\nabla^k \left\{ \frac{\left(\nabla_i \nabla_k - g_{ik}\nabla_m
\nabla^m \right){\cal F}(\Phi) +   T^{(YM)}_{ik} +
T^{(\Phi)}_{ik} + T^{({\rm NonMin})}_{ik} }{1+\kappa {\cal
F}(\Phi) } \right\} = 0
\label{Eeeq}
\end{equation}
is satisfied identically, when $F^{(a)}_{ik}$ is a solution of the
Yang-Mills equations (\ref{Heqs}), and $\Phi$ is the solution of
(\ref{Heq}). Thus, we deal with self-consistent system of master
equations (\ref{Heqs}), (\ref{Heq}) and (\ref{Ein})- (\ref{TV}),
which form the non-minimally extended Einstein-Yang-Mills-dilaton
model with eight arbitrary functions ${\cal F}_0(\Phi)$, ${\cal
F}(\Phi)$, $f_0(\Phi)$, $f_1(\Phi)$, ... $f_5(\Phi)$.

\section{Application of the non-minimal EYMd theory \\
to the model with pp-wave symmetry}

\subsection{Reduction of master equations}

Let us consider now a plane-symmetric space-time associated
usually with a gravitational radiation. We assume the metric to be
of the form
\begin{equation}
ds^2 = 2dudv -L^2(u) \left[e^{2\beta(u)} (dx^2)^2 + e^{-2\beta(u)}
(dx^3)^2 \right]
 \,, \label{pp1}
\end{equation}
where $u= (t-x^1)/\sqrt2$ and $v= (t+x^1)/\sqrt2$ are the retarded
and advanced time, respectively. This space-time is known to admit
the $G_5$ group of isometries \cite{ExactSolution}, and three
Killing four-vectors, $\xi^k$, $\xi^k_{(2)}$ and $\xi^k_{(3)}$
form three-dimensional Abelian subgroup $G_3$. The four-vector
$\xi^k$ is the null one and covariantly  constant, i.e.,
\begin{equation}
\xi^k = \delta^k_{v} \,, \quad g_{kl} \ \xi^k \xi^l = 0 \,, \quad
\nabla_l \ \xi^k =0 \,. \label{pp2}
\end{equation}
The four-vectors $\xi^k_{(\alpha)}$ ($\alpha = 2,3$) are
space-like and orthogonal to $\xi^k$ and to each other, i.e.,
\begin{equation}
\xi^k_{(\alpha)} = \delta^k_{\alpha} \,, \quad g_{kl} \
\xi^k_{(2)} \xi^l_{(3)} = 0 \,, \quad g_{kl} \ \xi^k
\xi^l_{(\alpha)} = 0 \,. \label{pp3}
\end{equation}
The non-vanishing components of the Ricci and Riemann tensors are,
respectively
\begin{equation}
R_{uu}=R^2_{ \ u2u}+R^3_{ \ u3u} \,, \quad R^2_{ \ u2u} = -
\left[\frac{L^{\prime \prime}}{L} + (\beta^{\prime})^2 \right] -
\left[2\beta^{\prime} \ \frac{L^{\prime}}{L} +  \beta^{\prime
\prime} \right] \,, \label{ppRie1}
\end{equation}
\begin{equation}
R^3_{ \ u3u} = - \left[\frac{L^{\prime \prime}}{L} +
(\beta^{\prime})^2 \right] + \left[2\beta^{\prime} \
\frac{L^{\prime}}{L} +  \beta^{\prime \prime} \right] \,, \quad
R=0 \,. \label{ppRie2}
\end{equation}
We consider a {\it toy-model}, which satisfies the following
requirements. {\it First}, the potentials of the Yang-Mills field
are parallel in the group space \cite{Yasskin}, i.e.,
\begin{equation}
A^{(a)}_k = q^{(a)} A_k \,,  \quad G_{(a)(b)} q^{(a)}q^{(b)} = 1
\,, \quad q^{(a)} = const \,. \label{pp5}
\end{equation}
{\it Second}, the vector field $A_k$ and scalar field $\Phi$
inherit the symmetry of the space-time, i.e., the Lie derivatives
of these quantities along generators of the group $G_3$, $\{\xi \}
\equiv \{ \xi^k, \xi^k_{(2)}, \xi^k_{(3)} \}$, are equal to zero:
\begin{equation}
\pounds_{\{\xi\}} A_k = 0 \,, \quad \pounds_{\{\xi\}} \Phi = 0 \,.
\label{pp6}
\end{equation}
{\it Third}, the vector field $A^k$ is transverse, i.e.,
\begin{equation}
A^k = - \left[ A_{2} \xi^k_{(2)} + A_{3} \xi^k_{(3)} \right] \,,
\quad \xi^k A_k =0 \,. \label{pp7}
\end{equation}
{\it Fourth}, the potential of the scalar field is equal to zero,
$V(\Phi)=0$. {\it Fifth}, the cosmological constant is absent,
$\Lambda=0$. These five requirements lead to the following
simplifications.

\noindent {\it (i)} The fields $A_k$ and $\Phi$ depend on the
retarded time $u$ only; there are two non-vanishing components of
the field strength tensor
\begin{equation}
F^{(a)ik} = q^{(a)} \left[ \left( \xi^i \xi^k_{(2)} - \xi^k
\xi^i_{(2)} \right) A^{\prime}_{2}(u) + \left(\xi^i \xi^k_{(3)} -
\xi^k \xi^i_{(3)}\right) A^{\prime}_{3}(u) \right] \,,
\end{equation}
the invariant $F^{(a)}_{ik}F_{(a)}^{ik}$ as well as the terms
${\cal R}^{ikmn}F_{mn}^{(a)}$ are equal to zero.

\noindent {\it (ii)} The equations (\ref{Heqs}) and (\ref{Heq})
are satisfied identically.

\noindent {\it (iii)} The non-minimal terms $T^{(I)}_{ik}$, ...,
$T^{(V)}_{ik}$ (\ref{TI})-(\ref{TV}) disappear, and the functions
$f_1(\Phi)$, $f_2(\Phi)$,..., $f_5(\Phi)$, being non-vanishing,
happen to be hidden, i.e., they do not enter the equations for the
gravity field.

\noindent After such simplifications the non-minimal equations
for the gravity field (\ref{Ein})-(\ref{TV}) reduce to one
equation
\begin{equation}
\frac{L^{\prime \prime}}{L} + (\beta^{\prime})^2 =- \frac{1}{2}
\kappa T(u)\,, \label{pp8}
\end{equation}
where
\begin{equation}
T(u) {=} \left\{\frac{f_0(\Phi)}{L^2} \left[e^{{-}2\beta}
\left(A^{\prime}_{2}\right)^2 {+} e^{2\beta}
\left(A^{\prime}_{3}\right)^2\right] {+}
\left(\Phi^{\prime}\right)^2 \left[ {\cal F}_0(\Phi){+}
\frac{d^2}{d\Phi^2} {\cal F}(\Phi)\right] {+} \Phi^{\prime \prime}
\frac{d}{d\Phi} {\cal F}(\Phi) \right\}\left[1 {+} \kappa {\cal
F}(\Phi)\right]^{{-}1} \,. \label{pp9}
\end{equation}
For this model $A_2(u)$, $A_3(u)$, $\Phi(u)$ are arbitrary
functions of the retarded time, and the prime denotes the
derivative with respect to $u$.

\subsection{Regular solutions}

Let us consider a  model with $L(u) \equiv 1$. It can be indicated
as a regular model, since $det(g_{ik})= -L^4 \equiv -1$ and can
not vanish, contrary to the standard situation with gravitational
pp-waves \cite{MTW}. When ${\cal F} = 0$, and ${\cal F}_0(\Phi)$,
$f_0(\Phi)$ are positive functions, there are no real solutions of
the equation (\ref{pp8}) with $L=1$, since the right-hand side is
negative for arbitrary moment of the retarded time (see
(\ref{pp9})). Nevertheless, such a possibility appears in the
non-minimal case. Below we consider two explicit examples of the
exact regular solutions.

\subsubsection{First explicit example}

Let the scalar field take a constant value $\Phi_0 \neq 0$, then
the functions
\begin{equation} \label{e1}
 A_2(u)= A_2(0) \ e^{\beta(u)} \,, \quad
A_3(u)= A_3(0) \ e^{- \beta(u)} \,,
\end{equation}
give the exact solution of (\ref{pp8}) for arbitrary
$\beta^{\prime}(u)$, when
\begin{equation}
- \kappa {\cal F}(\Phi_0) = 1 + \frac{\kappa}{2} f_0(\Phi_0)\left[
 A^2_2(0) + A^2_3(0)\right] \,. \label{e243}
\end{equation}
For a given negative function ${\cal F}(\Phi)$ and positive
function  $f_0(\Phi)$, this equality predetermines some special
value of the dilaton field, $\Phi_0$. The function $\beta(u)$ is
now arbitrary, and we can use, for instance, the periodic finite
function
\begin{equation}
\beta(u) = \frac{1}{2} h \ (1-\cos{2\lambda u}) \,, \quad \beta(0)
= 0 \,, \quad \beta^{\prime}(0) = 0 \,. \label{e343}
\end{equation}
The metric for this non-minimal model is periodic and regular
\begin{equation}
ds^{2} = 2 du dv - \left[e^{ 2 h \sin^2{\lambda u}} (dx^2)^2 +
e^{-2 h \sin^2{\lambda u}} (dx^3)^2 \right] \,, \label{30}
\end{equation}
the potentials of the Yang-Mills field (\ref{e1}) are also
periodic and regular.

\subsubsection{Second explicit example}

Let the dilaton field be periodic, i.e.,
\begin{equation}
\Phi(u) = \Phi_0 \sin{\lambda u} \,, \label{d1}
\end{equation}
and the color wave be circularly polarized, i.e.,
\begin{equation}
 A^{\prime}_2(u)= E_0 \  e^{\beta(u)} \cos{\omega u} \,, \quad
A^{\prime}_3(u)= E_0 \  e^{-\beta(u)} \sin{\omega u} \,.
\label{d2}
\end{equation}
The wave is called circularly polarized in analogy with
electrodynamics, since the function
\begin{equation}
 E^2(u) \equiv - g^{ik} A^{\prime}_i  A^{\prime}_k =
e^{{-}2\beta} \left(A^{\prime}_{2}\right)^2 {+} e^{2\beta}
\left(A^{\prime}_{3}\right)^2 = E^2_0  \label{d3}
\end{equation}
is constant for such a field. Let us consider the functions ${\cal
F}_0(\Phi)$, $f_0(\Phi)$ and ${\cal F}(\Phi)$ to be of the form
\begin{equation}
{\cal F}_0(\Phi) = 1 \,, \quad f_0(\Phi) = 1 +
\left(\frac{\Phi(u)}{\Phi_0}\right)^2  \,, \quad {\cal F}(\Phi) =
- \left(1+ \frac{2}{\kappa \Phi^2_0} \right) \Phi^2(u) \,,
\label{d5}
\end{equation}
and assume for simplicity, that $E_0 = \Phi_0 \lambda$. Then the
equations (\ref{pp8}) and (\ref{pp9}) yield
\begin{equation}
\left(\beta^{\prime}(u) \right)^2 = 2 \lambda^2 \,, \qquad
\beta(u) = \pm \sqrt2 \lambda u \,. \label{d6}
\end{equation}
Thus, the metric takes the form
\begin{equation}
ds^{2} = 2du dv - \left[e^{\pm 2\sqrt2 \lambda u} (dx^2)^2 +
e^{\mp 2\sqrt2 \lambda u} (dx^3)^2 \right] \,, \label{d7}
\end{equation}
and the space-time is symmetric \cite{ExactSolution}, i.e., all
the non-vanishing components of the Riemann tensor are constant
\begin{equation}
R^2_{ \ u2u} = R^3_{ \ u3u} = - (\beta^{\prime})^2 = - 2
\lambda^2  = \frac{1}{2} R_{uu} \,. \label{Petrov}
\end{equation}
The Yang-Mills potentials are quasi-periodic functions
\begin{equation}
A_2(u) = \frac{E}{(\omega^2+2\lambda^2)} \left[e^{\pm \sqrt2
\lambda u}(\sqrt2 \lambda \cos{\omega u} + \omega \sin{\omega u})
- \sqrt2 \lambda \right] \,,  \quad A_3(0) = 0 \,, \label{d8}
\end{equation}
\begin{equation}
A_3(u) = \frac{E}{(\omega^2+2\lambda^2)} \left[e^{\mp \sqrt2
\lambda u}(\sqrt2 \lambda \sin{\omega u} - \omega \cos{\omega u})+
\omega\right] \,, \quad A_3(0) = 0 \,. \label{d81}
\end{equation}
This solution is also free of singularity at the finite moments of
the retarded time.

\section{Conclusions}

The Lagrangian of the presented non-minimal
Einstein-Yang-Mills-dilaton model includes eight arbitrary
functions depending on the dilaton field, thus, we deal with a
wide freedom of modeling. The first (simplest) example of the
application shows, that for the model with pp-wave symmetry five
of the eight arbitrary functions happen to be hidden, i.e., they
do not enter the master equations. Nevertheless, the presence of
three remained functions of the dilatonic field allows us to find
exact explicit solutions of the whole self-consistent system of
master equations, which can be indicated as regular
(quasi)periodic solutions of the pp-wave type. This means, that
the dilatonic extension of the non-minimal field theory seems to
be a promising instrument for its modification.

We assume, that this non-minimal EYMd model might be fruitful for
cosmological applications, especially, for the explanation of the
accelerated expansion of the Universe and dark energy phenomenon.
We also believe, that the mentioned wide freedom of modeling in
the framework of the non-minimal EYMd theory can be used in
searching for new regular solutions, describing colored  static
spherically symmetric objects. We intend to discuss these
problems in detail in further papers.

\vspace{5mm} \noindent {\bf Acknowledgments}

\noindent The authors are grateful to Professor P.M. Lavrov for
the invitation to contribute to this special issue. This work was
supported by the DFG through the project No. 436RUS113/487/0-5.

\end{document}